\documentclass[review]{elsarticle}

\usepackage{url}
\usepackage{amssymb}
\usepackage{amsmath}
\usepackage{graphicx}
\usepackage{color}
\usepackage{float}
\usepackage{tabularx}
\usepackage{array}
\usepackage[table]{xcolor}
\usepackage{lineno,hyperref}
\usepackage{multirow}
\usepackage{booktabs}
\usepackage{xcolor,colortbl}
\usepackage{enumitem}
\usepackage{fancyhdr}

\definecolor{grey}{rgb}{0.8,0.8,0.8}

\modulolinenumbers[5]

\journal{Journal of Information Security and Applications}

\floatstyle{boxed}
\newfloat{scheme}{ht}{lo}
\floatname{scheme}{Protocol}

\newcommand{\M}{\text{MOA-ID}}
\newcommand{\A}{\text{SRA}}

    \fancypagestyle{pprintTitle}{%
    \lhead{\scriptsize Bernd Zwattendorfer, Daniel Slamanig, The Austrian eID ecosystem in the public cloud: How to obtain privacy while preserving practicality, Journal of Information Security and Applications, Available online 8 December 2015, ISSN 2214-2126, http://dx.doi.org/10.1016/j.jisa.2015.11.004.
(http://www.sciencedirect.com/science/article/pii/S2214212615000642)
} \chead{}\rhead{}
    \lfoot{\emph{Preprint submitted to Journal of Information Security and Applications October 1, 2015} \\
		\textcopyright 2015 This manuscript version is made available under the CC-BY-NC-ND 4.0 license 
http://creativecommons.org/licenses/by-nc-nd/4.0/}\cfoot{}\rfoot{}
    
    }

\begin{document}

\begin{frontmatter}

\title{The Austrian eID Ecosystem in the Public Cloud: How to Obtain Privacy While Preserving Practicality}

%

\author[mymainaddress]{Bernd Zwattendorfer\corref{mycorrespondingauthor}}
\ead{bernd.zwattendorfer@iaik.tugraz.at}
%
\author[mymainaddress]{Daniel Slamanig}
\cortext[mycorrespondingauthor]{Corresponding author}
\address[mymainaddress]{Graz University of Technology, Inffeldgasse 16a, 8010 Graz, Austria}

\begin{abstract}
The Austrian eID system constitutes a main pillar within the Austrian e-Government strategy. The eID system ensures unique identification and secure authentication for citizens protecting access to applications where sensitive and personal data is involved. In particular, the Austrian eID system supports three main use cases: Identification and authentication of Austrian citizens, electronic representation, and foreign citizen authentication at Austrian public sector applications. For supporting all these use cases, several components -- either locally deployed in the applications' domain or centrally deployed -- need to communicate with each other. While local deployments have some advantages in terms of scalability, still a central deployment of all involved components would be advantageous, e.g. due to less maintenance efforts. However, a central deployment can easily lead to load bottlenecks because theoretically the whole Austrian population as well as -- for foreign citizens -- the whole EU population could use the provided services. To mitigate the issue on scalability, in this paper we propose the migration of main components of the ecosystem into a public cloud. However, a move of trusted services into a public cloud brings up new obstacles, particular with respect to privacy. To bypass the issue on privacy, in this paper we propose an approach on how the complete Austrian eID ecosystem can be moved into a public cloud in a privacy-preserving manner by applying selected cryptographic technologies (in particular using proxy re-encryption and redactable signatures). Applying this approach, no sensitive data will be disclosed to a public cloud provider by still supporting all three main eID system use cases. We finally discuss our approach based on selected criteria.
\end{abstract}


\end{frontmatter}
\thispagestyle{fancy}


\section{Introduction}

Unique identification and secure authentication are essential processes especially in security-sensitive areas of application such as e-Government or e-Health. In particular, these processes play a key role if sensitive data is processed. To ensure a high level of security for citizen applications in these areas, many European countries have already rolled out national eID solutions supporting unique identification and secure authentication. In Austria, the Austrian citizen card is the official eID for citizens \cite{Leitold2002}. 

In general, the Austrian e-Government strategy foresees a thorough eID concept based on the Austrian citizen card, which constitutes the core component for secure identification and authentication of citizens at Austrian e-Government applications. Moreover, the Austrian eID concept also contains representative authentications and authentications of foreign EU citizens, which are treated equally to Austrian citizens in e-Government scenarios. Hence, the main functions of the Austrian eID system are Austrian citizen identification and authentication at online applications, citizen authentication on behalf of a natural or legal person, and the support of foreign citizen authentication at Austrian e-Government applications.

To make these main functions work, the Austrian Austrian eID system involves several other components -- besides the Austrian citizen card -- which are interconnected to each other. Key components, amongst others, are for instance MOA-ID (Module for Online Applications -- Identification) \cite{Lenz2014}, an open source software component locally deployed in each service providers domain facilitating citizen card access, the MIS (Management Issuing Service) \cite{Leitold2011}, which constitutes a central service issuing electronic mandates, or the SPR-GW (SourcePIN Register-Gateway) \cite{Lenz2015}, which a central gateway supporting registration of foreign citizens in Austrian national population registers. Details on the individual components will be given in Section \ref{sec:TheAustrianEIDArchitecture}. Currently, the Austrian eID system treats several deployed MOA-ID instances as well as the MIS and the SPR-GW as trusted entities. While the local deployment model has indeed some benefits, particularly with respect to scalability, in some situations a centralized deployment approach -- besides the MIS and the SPR-GW -- also of MOA-ID may be preferable. However, in terms of scalability (theoretically the whole Austrian population could use these central services for identification and authentication at service providers) the existing approaches may reach their limits. This can easily lead to load bottlenecks at MOA-ID, the MIS, or the SPR-GW. While the use of electronic mandates and foreign citizen authentications are still in its start-up phase, frequent usages are to be expected in the future. The use of electronic mandates in Austria gets increasing popularity. For instance, professional representation or natural-to-legal person representation constitute daily business in legal procedures. Additionally, representation of parents for their children or children for elderly people are frequent use cases especially in health services. Furthermore, cross-border identifications are steadily increasing because the European Commission currently heavily pushes the STORK framework \cite{Leitold2011b}, which will be probably the dominant authentication framework across Europe in the future.

Coping with such increased load may not be easy to handle within the current central deployment scenarios, where each entity is deployed in a trusted data center. Therefore, the authors propose a move of important components of the Austrian eID system (e.g., MOA-ID, MIS, SPR-GW) into a public cloud. Deployment in a public cloud could definitely mitigate any scalability issues due to the characteristics (high scalability, high elasticity, cost reduction, etc.) provided by a public cloud environment. However, a move of such trusted service into a public cloud brings up new obstacles, particularly with respect to citizen’s privacy \cite{Pearson2010,Zissis2012,Sen2013}. Although privacy and security are one of the main issues of public clouds, we still consider the public cloud as the most promising cloud deployment model for a migration of governmental services such as the Austrian eID infrastructure into the cloud. The reasons are -- amongst others -- particularly the ability to absorb unforeseeable load peaks almost seamlessly and its huge cost savings potential compared to other cloud deployment models \cite{Harms2010, Zwattendorfer2013PublicCloudeGovernment}. While privacy in the current scenarios is ensured through organizational means, in this paper we illustrate how such a move of trusted services of the Austrian eID system into a public cloud can be successfully realized using cryptographic technologies (by particularly using proxy re-encryption and redactable signatures) by still preserving citizens’ privacy. 

The paper is structured as follows. Section \ref{sec:RelatedWork} briefly explains related work in the context of identity management. Cryptographic building blocks our work is based on are described in Section \ref{sec:CryptographicBuildingBlocks}. In Section \ref{sec:TheAustrianEIDConcept} the Austrian eID system and its individual components are described in detail. In addition, the three main supported use cases (identification and authentication of Austrian citizens, in representation, and of foreign citizens) and corresponding process flows are explained. How the individual components can be moved into a public cloud in a privacy-preserving manner and how the process flows will change is elaborated in Section \ref{sec:PortingTheAustrianEIDArchitectureIntoThePublicCloud}. In Section \ref{sec:Discussion} we discuss our approach with respect to security, privacy, and practicability. Finally, we draw conclusions in Section \ref{sec:Conclusions}.

\section{Related Work}
\label{sec:RelatedWork}

Identity management is no new topic and thus several identity management solutions exist. In this section we briefly outline a couple of identity management systems that have evolved over the past years \cite{Bauer2005a, Dabrowski2008d, Cao2010, Ferdous2012}.

First systems arose due to the need of managing employee's accounts in single organizations. User and identity data was simply stored in directories such as LDAP (Lightweight Directory Access Protocol). In this case, the scope of the identity management system was limited to this single organization. 

Since the need for cross-organizational communication and hence exchanging identification and authentication data across domains gained importance, more sophisticated identity management solutions have established. One early example of such systems is Kerberos \cite{rfc4120}, which enables secure and uniform authentication in insecure TCP/IP networks. While additionally the WWW became increasingly popular at this time, identity management systems on application level arose. 

One example for a central identity management system was Microsoft Passport (latterly called Windows Live ID\footnote{\url{https://login.live.com}}). Other systems, which follow a decentralized and federated architecture, were the Liberty Alliance Project\footnote{\url{http://www.projectliberty.org}} (that evolved to the Kantara initiative\footnote{\url{http://kantarainitiative.org}}) or Shibboleth\footnote{\url{http://shibboleth.net}}. Both projects, Liberty Alliance and Shibboleth, influenced the development of the current version of the Security Assertion Markup Language (SAML 2.0)\cite{SAML2_Tech_Overview}. SAML defines one of the most important standards dealing with Single Sign-On (SSO) and identity federation at the present time. Contrary to SAML, which is XML-based, OpenID\footnote{\url{http://openid.net/specs/openid-authentication-2_0.html}} or OpenID Connect\footnote{\url{http://openid.net/connect/}} rely on more light-weight protocols or data structures for identity data exchange, e.g., simple URL parameters or JSON Tokens. However, both OpenID and OpenID Connect model similar use cases like SAML. Other systems, which are deployed in the field but gained less importance so far are, e.g., WS-Federation\cite{WS-Fed}, Windows Cardspace\footnote{\url{http://msdn.microsoft.com/en-us/library/aa480189.aspx}}, or the Central Authentication Service (CAS)\footnote{\url{http://www.jasig.org/cas}}.

Unique identification and secure authentication are also essential in sensitive areas of application, e.g., in e-Government or e-Business. Many European countries have already rolled-out national eID solutions to their citizens, mostly based on smart cards or mobile phones. Examples of such national solutions are the German nPA \cite{Margraf2010}, the Belgian BELPIC \cite{Cock2004}, or the Austrian citizen card \cite{Leitold2002} (latter will be detailed in one of the next sections). Most national eID solutions rely on a Public Key Infrastructure (PKI) and X.509 certificates. The Modinis-IDM study \cite{DeCock2006}, the IDABC eID country reports \cite{IDABC_report}, or \cite{Arora2008, Arora2008a} give an extensive overview of national eID solutions in Europe.

Giving the emerging trend towards cloud computing, identity management gains also importance in this sector. Hence, different cloud identity models have already been defined to cover new requirements particularly relating to cloud computing \cite{Gopalakrishnan2009,Goulding2010,CloudSecurityAlliance2011,Cox2012,Zwattendorfer2014}. The most promising model to fully feature the cloud computing benefits is the operation of an identity provider in the cloud, mostly in the role of an identity broker \cite{Huang2010}. Examples for such implementations are Fugen's Cloud ID Broker\footnote{\url{http://fugensolutions.com/cloud-id-broker.
html}} or the SkIDentity project\footnote{\url{http://www.skidentity.com}}. However, those solutions totally neglect any privacy issues with respect to the cloud provider. 

To bypass this issue, a handful of privacy-preserving cloud identity management approaches have already emerged in the last years. For instance, Nunez et al. \cite{Nunez2012} proposed the integration of proxy re-encryption into the OpenID protocol. In follow-up work, they proposed a more generic privacy-preserving cloud identity management model, which they call \emph{BlindIdM} \cite{Nunez2014}. This model also applies proxy re-encryption but relies on SAML instead of OpenID for the transport protocol. A somewhat related architectural approach -- but particularly focusing on eIDs -- has been introduced in \cite{sacmat2014}. A completely different approach based on anonymous credentials for privacy-preservation has been proposed in \cite{Bjones2014}. 

Prior to this paper, in \cite{Zwattendorfer2015} we illustrate privacy-preserving design strategies for migrating the basic Austrian eID architecture into the public cloud. The three design strategies proposed there are based on proxy re-encryption, anonymous credentials, and fully homomorphic encryption respectively. Thereby, we conclude that using proxy re-encryption is the most practical approach. However, \cite{Zwattendorfer2015} only investigates the basic use case of the Austrian eID system, namely identification and authentication of Austrian citizens (see Section \ref{sec:TheAustrianEIDConcept} for details). In this paper we follow a similar approach using proxy re-encryption, but now illustrate the migration of the complete Austrian identity infrastructure into the public cloud. Thereby, we include the two other main uses cases (identification and authentication in representation and foreign citizen authentication), which in part have already been discussed previously in \cite{DBLP:conf/secrypt/ZwattendorferS13,DBLP:conf/trustcom/ZwattendorferS13}. However, we want to emphasize that it is not a simple combination of these existing results, but we aim at demonstrating that privacy-preserving identity management in public clouds using proxy re-encryption is also possible for complex systems such as the complete Austrian eID ecosystem, which has broad applicability.

\section{Cryptographic Building Blocks}
\label{sec:CryptographicBuildingBlocks}

Subsequently, we review cryptographic building blocks that are required within the proposed approach.

\subsection{Digital Signatures}
A digital signature scheme (DSS) is a triple $(\sf DSS.KG, DSS.Sign, DSS.Verify)$ of efficient algorithms, where $\sf DSS.KG$ is a probabilistic key generation algorithm that takes a security parameter $\kappa$ and outputs a private and public key pair $(sk,pk)$. The (probabilistic) signing algorithm $\sf DSS.Sign$ takes as input a message $m\in\{0,1\}^*$ and a private (signing) key $sk$, and outputs a signature $\sigma$. The verification algorithm $\sf DSS.Verify$ takes as input a public (verification) key $pk$, a message $m\in\{0,1\}^*$ and a signature $\sigma$, and outputs a single bit $b\in\{\tt true, false\}$ indicating whether $\sigma$ is a valid signature for $m$ under $pk$. One requires a DSS to be correct, i.e., all honestly generated signatures verify, and existentially unforgeable under adaptively chosen-message attacks (EUF-CMA). In practice one typically employs the hash-then-sign paradigm, i.e., instead of inputting $m$ into $\sf DSS.Sign$ and $\sf DSS.Verify$, one inputs ${\sf H}(m)$ where $\sf H$ is a suitable cryptographic hash function.

\subsection{(Public Key) Encryption}
A public key encryption (PKE) scheme is a triple $(\sf PKE.KG, PKE.Enc, PKE.Dec)$ of efficient algorithms, where $\sf PKE.KG$ is a probabilistic key generation algorithm that takes a security parameter $\kappa$ and outputs a private and public key pair $(sk,pk)$. The probabilistic encryption algorithm $\sf PKE.Enc$ takes as input a public key $pk$ and a message $m\in\{0,1\}^*$ and returns a ciphertext $c={\sf PKE.Enc}({pk}, m)$. The decryption algorithm $\sf PKE.Dec$ takes as input a private key $sk$ and a ciphertext $c$ and returns a message $m={\sf PKE.Dec}({sk},c)$ or $\bot$ in the case of failure. A PKE scheme needs to be correct, i.e., decrypting a ciphertext yields the encrypted message, and at least indistinguishable under chosen plaintext attacks (IND-CPA). 

Abstractly, we can define private key (or symmetric) encryption schemes (SE) analogously. $\sf SE.KG$ generates only a single key $k$ which is used as input to the encryption and decryption algorithms. For the security of a private key encryption scheme (SE) one also requires at least IND-CPA security. Note that when we speak of applying PKE to a message $m$, then we implicitly mean applying \emph{hybrid encryption}, i.e., generating a random key $k$ of an SE scheme and sending/storing the tuple $(c_1={\sf PKE.Enc}({k},{pk}),c_2={\sf SE.Enc}(m,{k}))$.

\subsection{Redactable Signatures}
A conventional DSS scheme does not allow for alterations of a signed message without invalidating the signature. So called malleable signatures allow to modify (specified) parts of a signed message without invalidating the signature. Malleable signature schemes which allow \emph{removal} of parts (replacement by some special symbol $\bot$) by \emph{any} party are called redactable signature (RS) schemes \cite{StBuZh01,JoMoXiWa02}. Basically, they can be constructed from any secure DSS relying on the hash-then-sign paradigm by virtue of modifying the construction of the hash value (typically using randomized Merkle-Hash trees instead of a plain cryptographic hash of the entire message). Besides RS constructions for linear documents, there are also approaches for tree-structured documents, e.g., XML documents \cite{DBLP:conf/acns/BrzuskaBDFFKMOPPS10,DBLP:conf/cms/SlamanigR10}. Below, we present an abstract definition of redactable signature schemes. Henceforth we assume that a secure scheme for linear documents is used and refer the reader to \cite{StBuZh01} for required security properties.
\begin{description}
  \item[\textsf{RS.KG}:] This probabilistic key generation algorithm takes a security parameter $\kappa$ and
  produces and outputs a private and public key pair $(sk,pk)$.
 \item[\textsf{RS.Sign}:] This (probabilistic) signing algorithm gets as input the signing key $sk$ and a message $m = (m[1],\ldots,m[\ell])$, split into blocks $m[i]\in \{0,1\}^*$, and outputs a signature $\sigma = {\sf RS.Sign}(sk,m)$.
 \item[\textsf{RS.Verify}:] This deterministic signature verification algorithm gets as input a public key $pk$, a message $m = (m[1],\ldots,m[\ell])$, $m[i]\in \{0,1\}^*$, and a signature $\sigma$ and outputs a single bit $b={\sf RS.Verify}(pk,m,\sigma)$, $b\in\{ \tt true, false\}$, indicating whether $\sigma$ is a valid signature for $m$ under $pk$.
 \item[\textsf{RS.Redact}:] This (probabilistic) redaction algorithm takes as input a message $m = (m[1],\ldots,m[\ell])$, $m[i]\in \{0,1\}^*$, a public key $pk$, a signature $\sigma$, and a list $\sf MOD$ of
indices of blocks to be redacted. It returns a modified message and signature pair $(\hat{m},\hat{\sigma})={\sf RS.Redact}(m,pk,\sigma,{\sf MOD})$ or an error. 
\end{description}
Note that for any redacted signature $(\hat{m},\hat{\sigma})$, we have that ${\sf RS.Verify}(pk,\hat{m},\hat{\sigma})=\tt true$ holds.

\subsection{Proxy Re-Encryption}
Proxy re-encryption (RE) \cite{DBLP:conf/eurocrypt/BlazeBS98} is a public key encryption paradigm where a semi-trusted proxy, given a transformation key, can transform a message encrypted under the key of party $A$ into another ciphertext to the same message such that another party $B$ can decrypt with its private key. Although the proxy can perform this re-encryption operation, it does not learn anything about the encrypted message. According to the direction of this re-encryption operation, such schemes can be classified into bidirectional, i.e., the proxy can transform from $A$ to $B$ and vice versa, and unidirectional, i.e., the proxy can convert in one direction only, schemes. Furthermore, one can distinguish between multi-use schemes, i.e., the ciphertext can be transformed from $A$ to
$B$ to $C$ etc., and single-use schemes, i.e., the ciphertext can be transformed only once. Moreover, it is desirable that an RE scheme is non-interactive, i.e., a transformation key from $A$ to $B$ can be locally computed by $A$, where only the public key of $B$ is required. In this approach we exemplary use the unidirectional multi-use identity-based proxy re-encryption scheme of Green and Ateniese \cite{DBLP:conf/acns/GreenA07}, as in our setting we have a master authority (SRA), which can take care of the key generation. For simplicity we omit the inclusion of the $\sf MaxLevels$ parameter (indicating the maximum number of re-encryptions) in our definitions below and note that this parameter needs to be adjusted as required.

\begin{description}
\item[\textsf{RE.Setup}:] This probabilistic algorithm gets a security parameter $\kappa$. It outputs the master public
parameters $params$, which are distributed to users, and the master private key $msk$, which is kept private. We assume that $params$ is available to all algorithms and do not explicitly indicate it.
\item[\textsf{RE.KG}:] This probabilistic key generation algorithm gets the master private key $msk$, and an identity $id\in \{0,1\}^*$ and outputs a private
key $sk_{id}$ corresponding to identity $id$.
\item[\textsf{RE.Enc}:] This probabilistic encryption algorithm gets an identity $id\in \{0,1\}^*$, and a plaintext $m$ and outputs $c_{id}={\sf RE.Enc}(id,m)$.
\item[\textsf{RE.RKGen}:] This probabilistic re-encryption key generation algorithm gets a private key $sk_{id_1}$  (derived via $\sf RE.KG$), and two identities $(id_1, id_2)\in(\{0,1\}^*)^2$ and outputs a re-encryption key $$rk_{id_1\rightarrow id_2}={\sf RE.RKGen}(sk_{id_1},id_1,id_2).$$
\item[\textsf{RE.ReEnc}:] This (probabilistic) re-encryption algorithm gets as input a ciphertext $c_{id_1}$ under identity $id_1$ and a re-encryption key
$rk_{id_1\rightarrow id_2}$ (generated by $\sf RE.RKGen$) and outputs a re-encrypted ciphertext $$c_{id_2}={\sf RE.ReEnc}(c_{id_1},rk_{id_1\rightarrow id_2}).$$
\item[\textsf{RE.Dec}:] This decryption algorithm gets a private key $sk_{id}$, and a ciphertext $c_{id}$ and outputs $m={\sf RE.Dec}(sk_{id},c_{id})$ or an error $\bot$.
\end{description}
We note that as with PKE schemes one requires at least IND-CPA security. 

\section{The Austrian eID Concept}
\label{sec:TheAustrianEIDConcept}

Unique identification is essential in sensitive areas of applications such as e-Government. Especially if the number of users increases, such as the population of a country, identification based on first name, last name, and date of birth may be ambitious. Therefore, in Austria all citizens are registered in the Central Register of Residence (CRR) and have a unique number assigned (CRR Number). However, this CRR number must not be used directly in e-Government applications due to legal data protection restrictions. Therefore, the SourcePIN Register Authority (SRA), a subdivision of the Austrian
Data Protection Commission, encrypts the CRR number to derive a new unique identifier, which is called sourcePIN. The sourcePIN is stored on the Austrian citizen card in conjunction with other identity data such as first name, last name, date of birth, and a qualified signing certificate bound to the citizen's identity. These identification data is wrapped in a special XML-based data structure, the so-called Identity Link. The Identity Link is electronically signed by the SourcePIN Register Authority, which ensures integrity and authenticity of the citizen's identification data on the one side, and, on the other side, certifies the link between identity data and the qualified signing certificate. The Identity Link is finally solely stored on the Austrian citizen card. To provide compact descriptions, we denote the Identity Link in a more general form by ${\cal I}=((A_1,a_1),\dots,(A_k,a_k))$ as a sequence of attribute labels and attribute values.

To preserve citizen's privacy, it is forbidden by law (based on the Austrian e-Government Act) to directly use the sourcePIN for identification at online applications. According to this act, the sourcePIN must also never be stored outside the Identity Link. Nevertheless, for still being able to uniquely identify Austrian citizens at applications, the Austrian e-Government concept and strategy foresees a sector-specific identification model. Thereby, a sector-specific PIN (ssPIN) is derived from the combination of the sourcePIN and a governmental sector identifier denoted as $s$ (e.g., finance, tax, etc.) by using cryptographic one-way hash functions. The use of one-way hash functions still ensures uniqueness of the identifier (ssPIN). In addition, it is not possible to either re-calculate the sourcePIN or an ssPIN of another sector out of a given ssPIN. The ssPIN is finally used for unique identification at online applications.

The entire Austrian eID concept for natural persons relies on citizens being registered in the CRR. However, persons not listed in the CRR (e.g., foreign citizens or Austrian citizens currently residing in a foreign country) can be registered in the so-called Supplementary Register for Natural Persons (SR). By registering in the SR, these persons also get a sourcePIN assigned and hence become part of the Austrian eID infrastructure. This way, foreign citizens can be treated equally to Austrian citizens in online applications. The legal basis for this treatment is the so-called E-Government Equivalence Decree, which became law in 2010.

Besides identification and authentication of natural persons (being either an Austrian or foreign citizen), the Austrian eID concept foresees electronic identification possibilities also for legal persons. Thereby, each legal person can be uniquely identified by a number which has been registered in one of several business registers for legal persons. Such registers are for instance the Company Register or the Central Register for Associations. In general, the overall identification process is based on the usage of electronic mandates. Electronic mandates can be used as electronic representations for legal persons, natural persons, or for professional representatives (e.g., lawyers, notaries, etc.). On a high level, for the representation of legal persons the unique number out of a business register and the representative's sourcePIN (natural person) form the basis information for an electronic mandate. In case of representation of two natural persons, the sourcePIN of both the representative and the empowering person (mandator) are taken for modelling the electronic mandate.

\subsection{The Austrian Citizen Card Concept}

The Austrian citizen card constitutes the official eID in Austria and is the key component within the Austrian eID concept. The Austrian citizen card concept is rather an abstract definition of a secure eID token than a concrete implementation. Due to this technology-agnostic concept different citizen card implementations exist. Current implementations in use are based on smart cards or mobile phones. However, due to the abstract definition and the technology-neutral concept also alternative approaches and implementations may be developed and rolled-out in the future.

Irrespective of the actual implementation, the citizen card provides the following functionality:

\begin{enumerate}
	\item Identification and authentication of Austrian citizens
	\item Qualified electronic signature creation
	\item Encryption and decryption.
	\item Data storage 	
\end{enumerate}

The Austrian citizen card can be used for uniquely \emph{identifying} citizens. Identification is based on the Identity Link, which is solely stored on the citizen card and which includes identifying information such as the sourcePIN. Since the sourcePIN cannot be used directly for identification at online applications because of data protection restrictions, it will be derived according to sectors which results in sector-specific PINs (ssPINs). These ssPINs are finally used for identification at online e-Government applications. \emph{Authentication} by using the Austrian citizen card is carried out by generating a \emph{qualified electronic signature}. The Austrian citizen card is capable for generating qualified electronic signatures according to the EU Signature Directive. Signatures created according to this directive are legally equivalent to handwritten signatures. However, this functionality is not only used for citizen authentication but also in other applications such as PDF document signing. The third functionality constitutes encryption and decryption. The citizen card includes an additional key pair which allows for secure hardware-based decryption of arbitrary data.\footnote{In the remainder of this paper, we denote the signature key pair of the citizen $C$ with $(pk_{C}', sk_{C}')$ and the encryption key pair with $(pk_{C}, sk_{C})$. For details on the formalism, we refer to Section \ref{sec:CryptographicBuildingBlocks}.} Finally, the third citizen card functionality is data storage, where data of arbitrary format (e.g. XML documents or digital certificates) can be stored on the card.

For accessing citizen card functionality irrespective of its implementation, an abstract access layer has been specified. This abstract layer hides implementation specifics from the application and enables access to citizen card functionality by using XML commands. Implementations of this abstract interface are called Citizen Card Software (CCS). The CCS can be either installed locally on the citizen's computer or is provided remotely on a server.

\subsection{The Austrian eID Architecture}
\label{sec:TheAustrianEIDArchitecture}

\begin{figure*}[ht!]
	\centering
		\includegraphics[width=13cm]{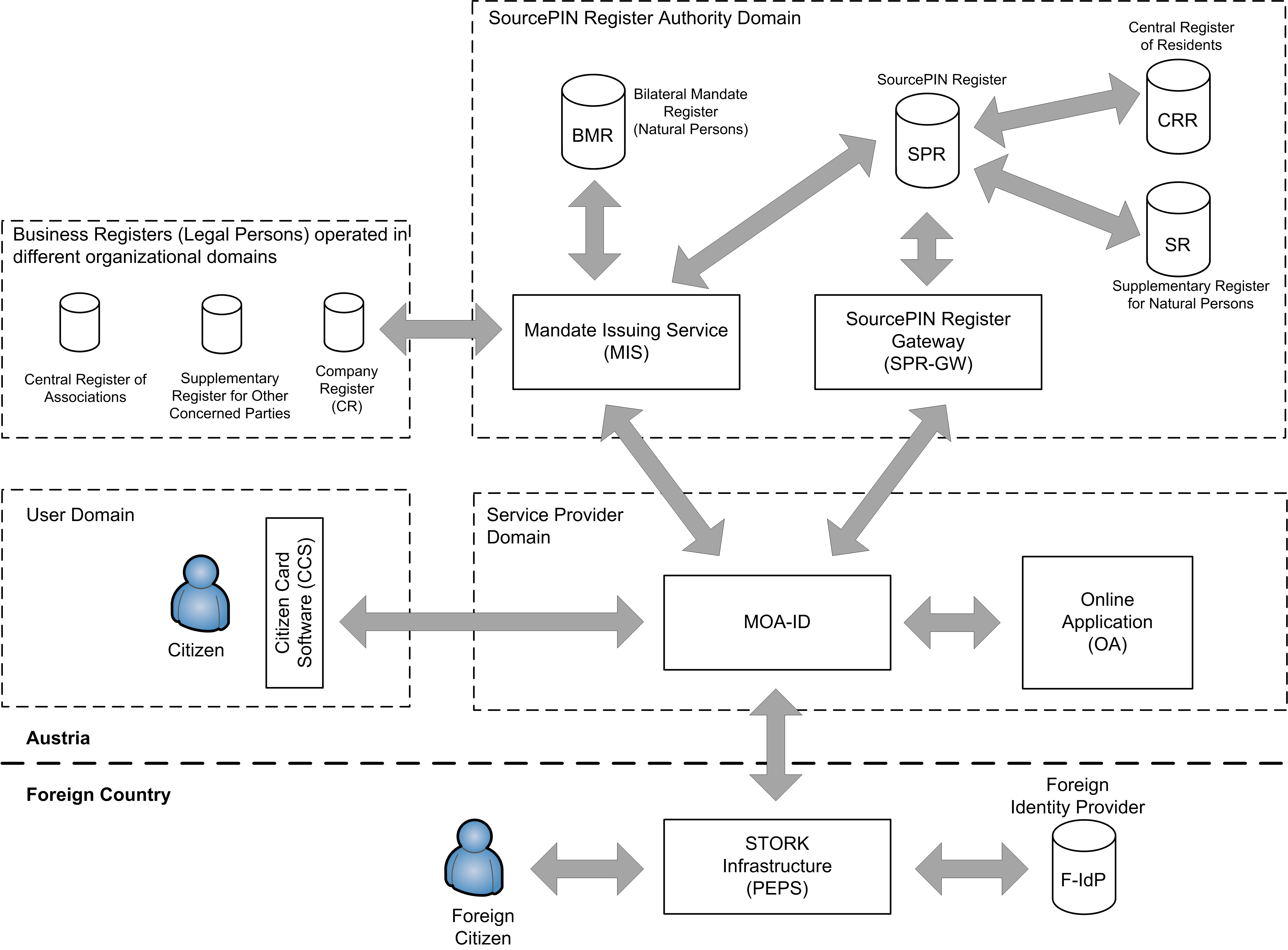}
	\caption{The Austrian eID Architecture}
	\label{fig:eID-Architecture}
\end{figure*}

The overall Austrian eID architecture involves several systems and components. Figure \ref{fig:eID-Architecture} illustrates the Austrian eID architecture separated into operational domains. In the following, we briefly describe the individual components and their basic functionality based on domain separation. Their interactions and individual process flows supporting different use cases will be described in the Section \ref{sec:IdentificationAndAuthenticationUseCases}. Details on the Austrian eID architecture can also be found in \cite{Stranacher2013}.

\begin{description}
	\item[User Domain:] A \emph{Citizen} wants to access public or private sector service using her Austrian citizen card. The \emph{Citizen Card Software}, which enables easy access to citizen card functionality, usually runs in the citizen's domain.
	\item[Service Provider Domain:] A service provider hosts one ore more public or private sector \emph{online applications} providing web-based services to citizens. These services require qualified and secure identification and authentication of the Austrian citizen card. Identification and authentication for the application is handled and managed by \emph{MOA-ID}. On the one hand, MOA-ID accesses citizen card functionality, and, on the other hand, provides specific and authentic citizen data to the online application for further processing.
	\item[SourcePIN Register Authority Domain:] The \emph{Mandate Issuing Service} (MIS) is only invoked if citizens authenticate as representative for a natural or legal person. The MIS issues electronic mandates on the fly. For querying appropriate mandate information for natural person representation, the MIS has to query the \emph{Bilateral Mandate Register} (BMR). For fetching appropriate mandate information for representing legal persons, an according \emph{Business Register} - depending on the type of the legal person - needs to be queried. To finish an authentication process using representation between natural persons, the \emph{SourcePIN Register} (SPR) needs to be queried. The SourcePIN Register is more ore less a virtual register, which bundles the information of the \emph{Central Register of Residents} (CRR) and the \emph{Supplementary Register for Natural Persons} (SR). The \emph{SourcePIN Register Gateway}, which is also operated within the SourcePIN Register Authority Domain, is only invoked in the case of foreign citizen authentication. Thereby, the SPR-GW facilitates the registration of foreign citizens in the SR for MOA-ID.
	\item[Business Registers:] In Figure \ref{fig:eID-Architecture} the individual business registers (\emph{Company
Register, Central Register of Associations, Supplementary Register for Other Concerned Parties}) are subsumed under one block for simplicity. However, the individual registers are actually operated in different organizational domains. Operators are for instance the Austrian Ministry of Justice or the Austrian Ministry of the Interior. These registers contain information of legal persons and hence also mandate information for their representation in electronic processes.
	\item[Foreign Country:] In most cases, foreign citizens authenticate via the \emph{STORK infrastructure}. The STORK infrastructure, operated in the foreign country, queries an appropriate \emph{Foreign Identity Provider (F-IdP)} for citizen identification and authentication. Authenticated citizen data are transferred via the STORK infrastructure into the Austrian eID system (more precisely to MOA-ID).
\end{description}

\subsection{Identification and Authentication Use Cases}
\label{sec:IdentificationAndAuthenticationUseCases}

The individual components work all together to support different use cases. In the following  we briefly describe three identification and authentication use cases. A detailed description of the interaction and communication between the individual components will be done in Section \ref{ia}.

\begin{description}
	\item[Identification and Authentication of Austrian Citizens:] For identification and authentication of Austrian citizens at online applications mainly the component MOA-ID is responsible. MOA-ID handles the identification process by reading and verifying the citizen's Identity Link, and by deriving the sector-specific PIN from the citizen's sourcePIN. Authentication is carried out by qualified signature creation, stating the willingness of authenticating at the online application. The citizen's signature is verified by MOA-ID. The complete identification and authentication process will be illustrated Figure \ref{fig:seq_aut_cloud}, illustrating also the equalities and differences between the current and the cloud process flow. The individual process steps are described in detail in Section \ref{ia_austrian_citizens}.
	\item[Identification and Authentication in Representation:] In addition to MOA-ID, in this scenario the Mandate Issuing Service (MIS) plays an important role. If a citizen wants to represent another person (natural or legal person) in an e-Government application, for successful authentication the citizen needs to provide an authentic electronic mandate to the online application. After the successful identification and authentication of an Austrian citizen, the citizen can select an electronic mandate via the MIS, which empowers her to represent the respective person. Details on this use case will be given in Section \ref{ia_mandate}. Figure \ref{fig:seq_mand_cloud} illustrates the identification and authentication scenario when representing a legal person currently and in the proposed cloud-based approach. For simplicity, we limit this use case and its description to legal person representation only, as the representation of natural persons is similar.
	\item[Identification and Authentication of Foreign Citizens:] The Austrian eID concept supports the secure identification and authentication of foreign citizens using their nationally-issued eID. In other words, foreign EU citizens can securely authenticate at Austrian online applications without having the need to apply for an Austrian eID but can use their own national one. For an online application a foreign citizen identification and authentication process is completely transparent, i.e. the foreign citizen can be treated equally to an Austrian citizen because the same citizen data and data format is used for transmission. For the support of foreign citizen identification and authentication, the Austrian eID architecture relies internally on the SPR-GW and on the external components provided by the foreign country (STORK infrastructure and F-IdP). Details on this use case will be given in Section \ref{ia_foreign_citizens} and the corresponding Figure \ref{fig:seq_foreign_cloud}.
\end{description}




\section{Porting the Austrian eID Architecture into the Public Cloud}
\label{sec:PortingTheAustrianEIDArchitectureIntoThePublicCloud}

As can be seen from Figure \ref{fig:eID-Architecture}, the individual components have different deployment approaches. For instance, MOA-ID follows a local deployment approach, where each service provider operates one MOA-ID instance in its domain. In comparison to that, the MIS and the SPR-GW are operated centrally in the domain of the SourcePIN Register Authority. Additionally, the deployment of the STORK infrastructure follows a central approach, where each Member State operates a central gateway (PEPS) providing cross-border eID functionality.

While the local deployment of MOA-ID has some clear advantages in terms of end-to-end security or scalability, a central approach may be still advantageous. Citizens could benefit from a central MOA-ID instance as they only need to trust one specific identity provider. Additionally, a central instance of MOA-ID would allow citizens single sign-on across different domains without re-authenticating at each service provider and online application respectively every time. Also service providers can benefit from such an approach as they do not require to run and maintain a separate MOA-ID instance. Naturally, a central deployment approach also has some drawbacks. Namely, a single instance constitutes a single point of failure or attack. Moreover, the level of scalability cannot be reached by a central approach compared to a local or distributed deployment.

Scalability is probably the main issue when considering a central deployment of MOA-ID as all citizen authentication processes will run through this central instance. This can easily lead to load bottlenecks, as theoretically the whole population of Austria could use this service. The same argument holds for the MIS, the SPR-GW, or the PEPS, which are currently all deployed centrally within a trusted environment. While the use of electronic mandates and cross-border authentications are still in its start-up phase, frequent usages are to be expected in the future. 

Dealing with such increased load may not be easy to handle within the current deployment scenarios, where each entity is deployed in a trusted data center. Therefore, we propose a move of the individual entities (MOA-ID, MIS, SPR-GW, PEPS) into a public cloud. Deployment in a public cloud could definitely mitigates any scalability issues due to the characteristics provided by a public cloud environment. However, a move of such trusted services into a public cloud brings up new obstacles, especially with respect to citizen's privacy. While privacy in the current scenarios is ensured through organizational means and mainly relies on trust, in the following sections we illustrate how such a move of trusted services into a public cloud can be successfully realized using cryptographic technologies by still preserving citizens' privacy.

The selected cryptographic technologies for the described approach are based on the results of a previous work \cite{Zwattendorfer2015}, which illustrates privacy-preserving design strategies for migrating the basic Austrian eID architecture into the public cloud only. In \cite{Zwattendorfer2015} the only use case that is analyzed is the identification and authentication of Austrian citizens. According to the results in \cite{Zwattendorfer2015}, the use of proxy re-encryption and redactable signatures is considered to yield the most practical approach. Thus, in this paper we continue our work by migrating the complete Austrian eID architecture (covering the additional use cases on electronic representation as well as foreign citizen identification and authentication) using proxy re-encryption and redactable signatures. While parts of the two other main uses cases (identification and authentication in representation and foreign citizen authentication) have previously been discussed in \cite{DBLP:conf/secrypt/ZwattendorferS13,DBLP:conf/trustcom/ZwattendorferS13}, in this paper we want to show the applicability of the resulting approach of \cite{Zwattendorfer2015} in a complex systems such as the complete Austrian eID ecosystem (with all required components interacting with each other), which has broad applicability.

\subsection{The Austrian eID Architecture in the Public Cloud}

\begin{figure*}[ht!]
	\centering
		\includegraphics[width=13cm]{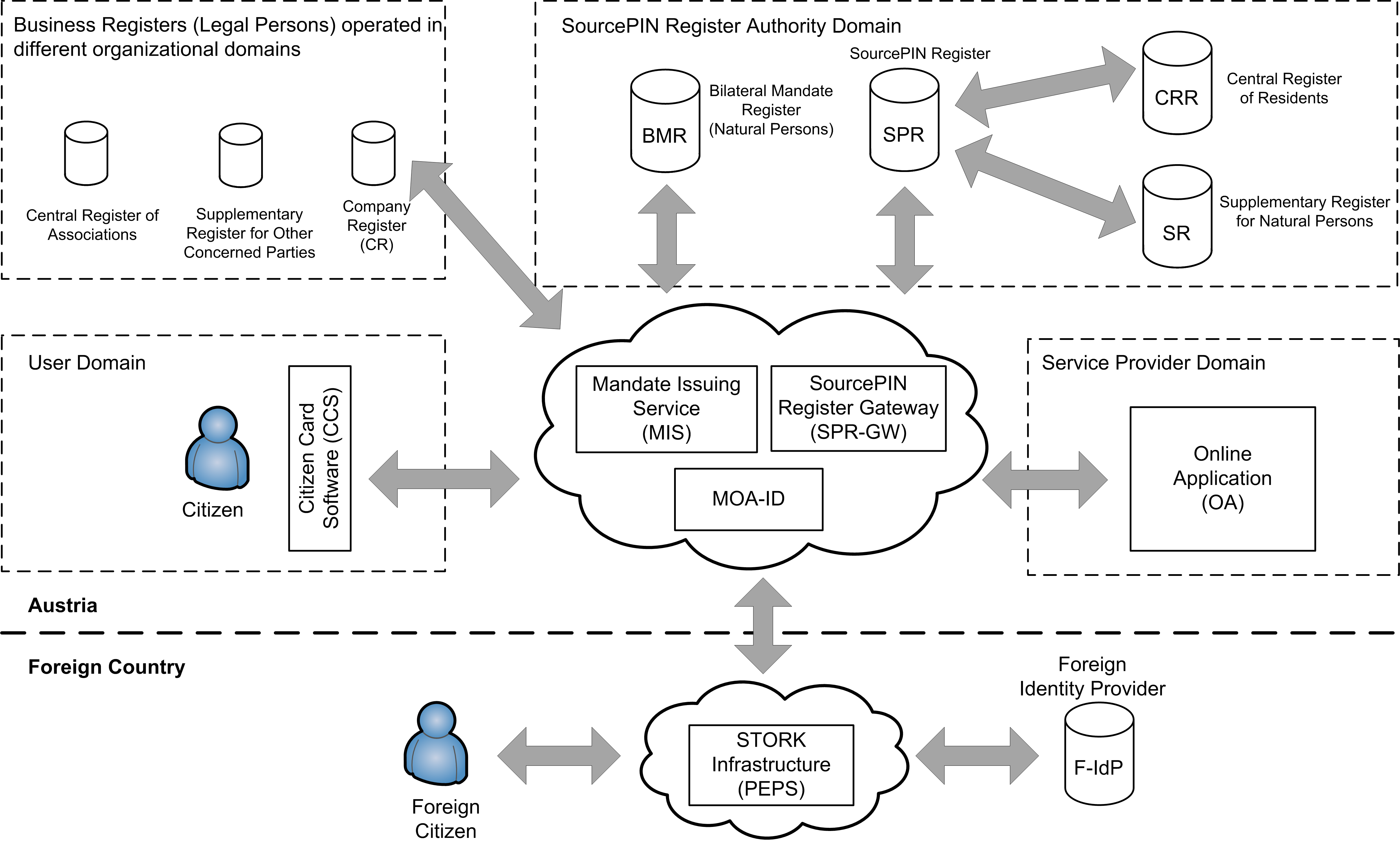}
	\caption{The Austrian eID Architecture in the Public Cloud}
	\label{fig:eID-Architecture-Cloud}
\end{figure*}

Figure \ref{fig:eID-Architecture-Cloud} illustrates the new architecture of the Austrian eID system when moving important components into the public cloud. In this figure, for simplicity we subsumed the components MOA-ID, MIS, and SPR-GW to be deployed in one public cloud. However, all three components could be operated by different public cloud providers. The STORK PEPS component is assumed to be operated in a different public cloud, as it will be under responsibility of the foreign country.

For being able to move the Austrian eID infrastructure into a public cloud, a few minor changes in the corresponding infrastructure are necessary. In the next sub-section, we explain in detail which changes are required. In the subsequent sections, we describe the adapted process flows of the individual use cases to support an operation of the Austrian eID system in a public cloud.

\subsection{Identification and Authentication Use Cases}
\label{ia}

\subsubsection{Identification and Authentication of Austrian Citizens}
\label{ia_austrian_citizens}

Basically, similar to the current situation we assume the SourcePIN Register Authority (SRA) as trusted entity. In this setup scenario, the SRA will be also responsible for the issuance of a slightly modified Identity Link. Additionally, the SRA will manage service provider registration to build appropriate trust relationships between the individual entities.

\paragraph{Setup}

In the proposed cloud scenario, we assume that the modified Identity Link (denoted by $\cal I'$) does not contain a sourcePIN but furthermore all ssPINs according to all governmental sectors. Furthermore, all ssPINs are encrypted using a proxy re-encryption scheme, hence every $(A_i,a_i)\in\cal I'$ is replaced by the SRA by the encrypted attributes $c_{a_i}={\sf RE.Enc}(\A,a_i)$ . The ssPINs and additional citizen attributes (e.g., name, date of birth) are encrypted under the public key of MOA-ID $(pk_{\M})$. The key pair $(pk_{\M},sk_{\M})$ is generated by the SRA. However, the SRA as trusted entity keeps the corresponding private key $(sk_{\M})$ and thus MOA-ID will not be able to decrypt the individual attributes. In the current approach, conventional signatures are used to ensure authenticity and integrity of the Identity Link. However, in this cloud-based approach the SRA signs the modified Identity Link using a redactable signature scheme resulting in $\sigma_{\cal I'}={\sf RS.Sign}(sk_{\A},{\cal I'})$. By this, each individual attribute of the modified Identity Link can be redacted. The modified Identity Link $\cal I'$ is finally stored on the citizen card. In this setup, we further assume that the signature creation certificate stored on the citizen card does not contain any citizen identifying information.

In addition, service providers need to register their online applications at the SRA. We denote the set of service providers $S=\{S_1,\dots,S_\ell\}$. For service provider registration, the SRA produces a private key $sk_{S_j}={\sf RE.KG}(msk_{\A},S_J)$ for $S_j$ and a re-encryption key $rk_{\M \rightarrow S_j}={\sf RE.RKGen}(sk_{\M},\M,S_j)$. The key $sk_{S_j}$ is issued to $S_j$ and $rk_{\M \rightarrow S_j}$ to MOA-ID. We further assume that an appropriate signing key pair $(pk_{\M}', sk_{\M}')$ for MOA-ID is available. 

\paragraph{Process Flow}

\begin{figure}[ht!]
	\centering
		\includegraphics[width=13cm]{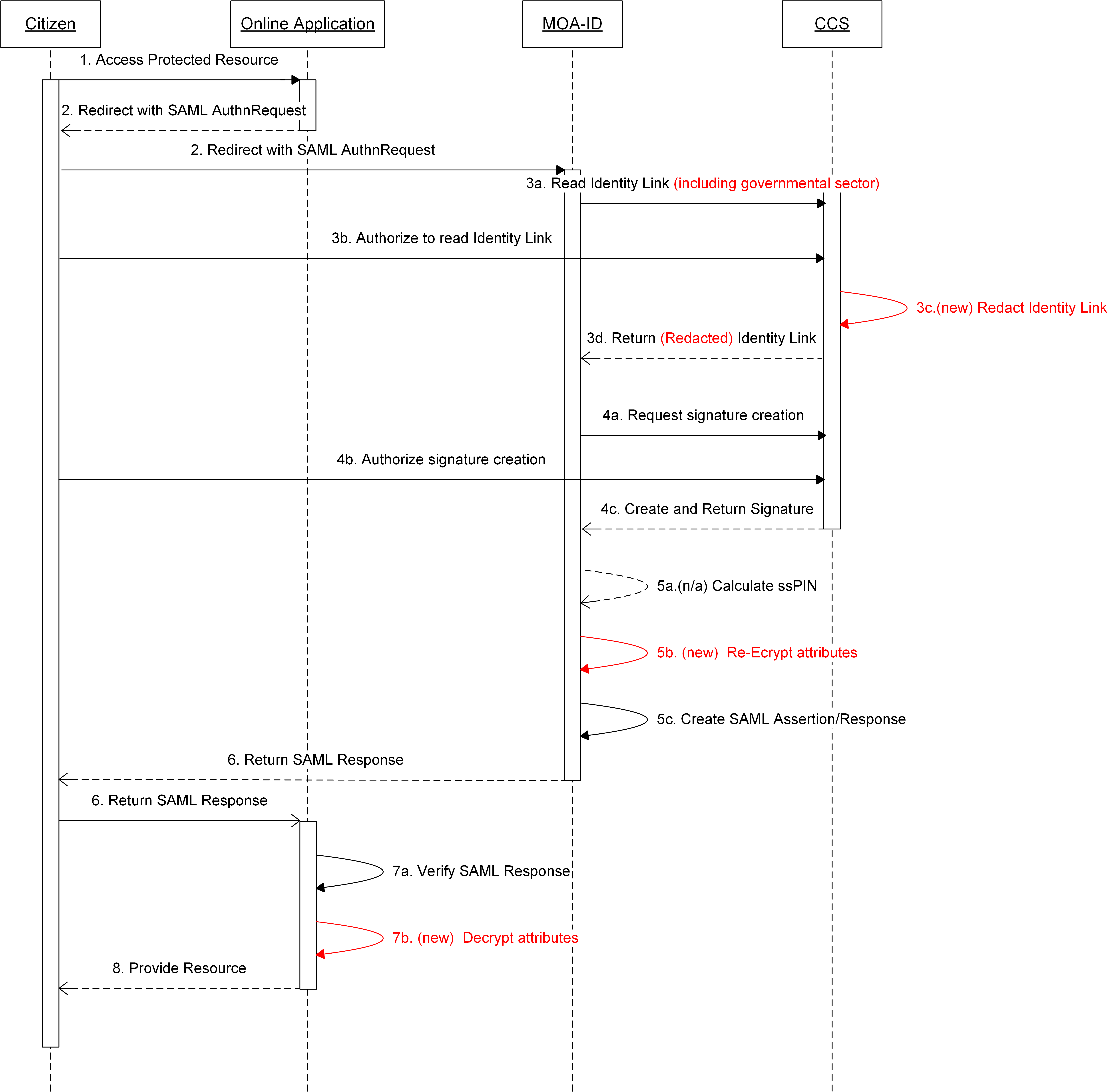}
	\caption{Process flow of Austrian citizen identification and authentication in the cloud approach}
	\label{fig:seq_aut_cloud}
\end{figure}

Figure \ref{fig:seq_aut_cloud} illustrates the process flow combining the current and the cloud-based approach. In fact, the cloud process flow is very similar to the current scenario. However, the differences in the cloud approach compared to the current approach are highlighted in red in Figure \ref{fig:seq_aut_cloud}. In the following, the current eID process flow as well as necessary modifications for the cloud deployment are described in detail.

\begin{enumerate}
	\item [1.] The citizen wants to access a protected resource at the online application, which requires proper authentication.
	\item [2.] The online application assembles a SAML authentication request, which is transmitted via HTTP-Redirect to MOA-ID.
	\item [3a.] In this step, MOA-ID sends an appropriate XML request to the CCS for retrieving the Identity Link from the citizen card. For the cloud approach, this request further includes now the governmental sector $s$. 
	\item [3b.] The citizen authorizes this request appropriately depending on the CCS implementation.
	\item [3c.(new)] By having $s$, the CCS can now redact all ssPINs except the ssPIN corresponding to $s$.
	\item [3d.] The identity link is returned to MOA-ID and verified. In the cloud approach, the redacted Identity Link $\cal I'$ is returned and verified.
	\item [4a.] MOA-ID requests the creation of a qualified electronic signature indicating the willingness of the citizen for online application authentication.
	\item [4b.] The citizen authorizes this request appropriately depending on the CCS implementation.
	\item [4c.] The citizen creates a signature, which is sent back to MOA-ID and verified.
	\item [5a.(n/a)] MOA-ID derives the appropriate ssPIN out of the sourcePIN for the sector the online application belongs to. This step is not applicable in the cloud approach as the sourcePIN as well as the ssPIN are valuable assets which must not to be disclosed to the cloud provider.
	\item [5b.(new)] Instead of deriving an ssPIN, MOA-ID re-encrypts the attributes $c_{a_i}$ of the redacted Identity Link $\cal I'$ for the authentication requesting service provider $S_j$ by using the re-encryption key $rk_{\M \rightarrow S_j}$. This results in $c_{S_j}={\sf RE.ReEnc}(rk_{\M \rightarrow {S_j}}, c_{a_i})$. 
	\item [5c.] In the current approach MOA-ID assembles a SAML assertion/response, which includes the ssPIN and additional citizen data out of the Identity Link. In the cloud approach, MOA-ID signs the result coming out with $\sigma_{\M}={\sf DSS.Sign}(sk_{\M}, c_{S_j})$. However, more precisely also in the cloud approach the complete SAML assertion/response is signed.
	\item [6.] MOA-ID returns the SAML assertion/response to the online application via HTTP-POST. Compared to the current approach, where attributes are included in plain in the SAML message, in the cloud approach the SAML message includes re-encrypted attributes only.
	\item [7a.] The online application verifies the SAML response ($\sigma_{\M}$), extracts its (enrypted) attributes.
	\item [7b.(new)] The encrypted citizen attributes $c_{S_j}$ are decrypted using the private key $sk_{S_j}$.	
	\item [8.] After successful verification, the online application grants access to the resource.
\end{enumerate}

\subsubsection{Identification and Authentication in Representation}
\label{ia_mandate}

Identification and authentication in representation requires a successful identification and authentication of an Austrian citizen as a prerequisite. After that, the Austrian citizen is eligible to select an electronic mandate containing necessary empowerment information for representing a natural or legal person. In the following, we first give details on the setup for supporting a migration of this use case into a cloud environment. In addition, we give details on the process flow highlight similarities and difference between the current process steps and the steps required in a cloud deployment.

\paragraph{Setup}
In this scenario, we again assume that the modified Identity Link $\cal I'$ is used. Furthermore, in this scenario we additionally rely on the encryption and decryption functionality of the Austrian citizen card. Besides a signature key pair, each Austrian citizen $C$ has an encryption key pair $(pk_{C},sk_{C})$ stored on her citizen card. This key pair is also generated by the SRA.

In addition to $(pk_{S_j}, sk_{S_j})$ and $rk_{\M \rightarrow S_j}$, the SRA has to generate additional encryption and re-encryption keys for the individual entities required for mandate processing. For the MIS and for the CR the keys $(pk_{MIS},sk_{MIS})$ and $(pk_{CR},sk_{CR})$ are created. Since the MIS will be operated in the cloud, the SRA keeps secret $sk_{MIS}$ and only distributes $pk_{MIS}$ to the MIS. In addition, the following re-encryption keys are generated: $rk_{\M \rightarrow {MIS}}$, $rk_{{MIS} \rightarrow {CR}}$, and $rk_{{MIS} \rightarrow \M}$. We further assume that appropriate signing keys are available for the individual entities: $(pk_{\M}', sk_{\M}')$, $(pk_{MIS}', sk_{MIS}')$, and $(pk_{CR}', sk_{CR}')$.

\paragraph{Process Flow}

\begin{figure}[ht!]
	\centering
		\includegraphics[width=13cm]{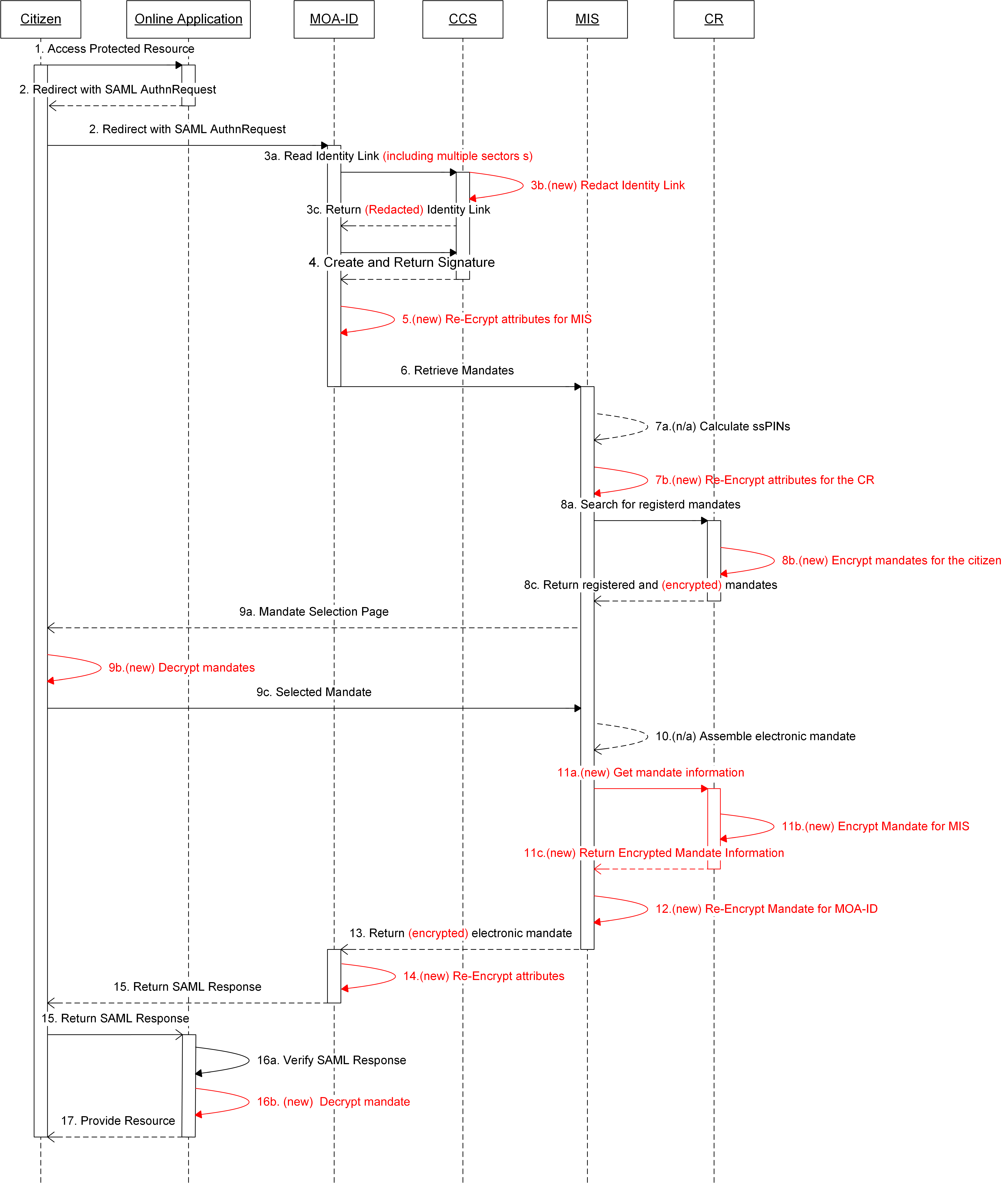}
	\caption{Process flow representing a legal person electronically in the cloud approach}
	\label{fig:seq_mand_cloud}
\end{figure}

Figure \ref{fig:seq_mand_cloud} illustrates the process flow for representative authentication in the current and in the cloud-based approach. In the following, we describe the process flow in detail.

\begin{enumerate}
	\item [1.] This process step is equal to normal and cloud-based Austrian citizen authentication (cf. Section \ref{ia_austrian_citizens}). However, the citizen indicates that she wants to authenticate on behalf of somebody (e.g., by activating a checkbox).
	\item [2.] This process step is equal to normal and cloud-based Austrian citizen authentication (cf. Section \ref{ia_austrian_citizens}).
	\item [3a.] In the current scenario, this process step is equal to normal Austrian citizen authentication (cf. Section \ref{ia_austrian_citizens}). For the cloud-based approach, this process step is only similar to cloud-based Austrian citizen authentication. When MOA-ID sends a request for retrieving the Identity Link to the CCS, the request includes the governmental sector the service provider belongs to and the governmental sectors the individual registers storing mandate information belong to. 
	\item [3b.(new)] Similar to the cloud-based Austrian citizen authentication approach, the CCS redacts all ssPINs which are not required in this authentication scenario. This includes all ssPINs except the one the service provider belongs to ($ssPIN_{SP}$) and the ssPINs required for querying the individual registers for mandate information ($ssPIN_{CR}$ in this example).
	\item [3c.] This process step is equal to normal and cloud-based Austrian citizen authentication (cf. Section \ref{ia_austrian_citizens}).
	\item [4.] This process step is equal to normal and cloud-based Austrian citizen authentication (cf. Section \ref{ia_austrian_citizens}).
	\item [5.(new)] In this step, MOA-ID re-encrypts the attribute $ssPIN_{CR}$ from $\cal I'$ for the MIS using $rk_{\M \rightarrow {MIS}}$ resulting in $c_{MIS}={\sf RE.ReEnc}(rk_{\M \rightarrow {MIS}}, ssPIN_{CR})$. This re-encryption result is signed by MOA-ID which outputs $\sigma_{MOA-ID}={\sf DSS.Sign}(sk_{\M}', c_{MIS})$.
	\item [6.] Since the citizen wants to authenticate on behalf of somebody, the MIS is queried by MOA-ID for accessing all mandates the citizen is empowered. For that, in the current approach MOA-ID sends the citizen's Identity Link to the MIS. 
	
	In the cloud-based approach MOA-ID sends the tuple ($c_{MIS}, \sigma_{\M}$) to the MIS for mandate retrieval.
	\item [7a.(n/a)] Out of the sourcePIN from the Identity Link, the MIS calculates all appropriate ssPINs for querying the individual registers. For simplicity, in this scenario the authors illustrate the query process at the company register (CR) only. This step is not applicable in the cloud approach as the sourcePIN as well as the ssPINs are valuable assets which must not to be disclosed to the cloud provider.
	\item [7b.(new)] The MIS verifies $\sigma_{\M}$ and re-encrypts $c_{MIS}$ for the CR using $rk_{{MIS} \rightarrow {CR}}$ and signs the result $c_{CR}$. The resulting signature is denoted as $\sigma_{MIS}={\sf DSS.Sign}(sk_{MIS}', c_{CR})$.
	\item [8a.] In the current approach, the MIS searches the CR for registered mandates using the corresponding $ssPIN_{CR}$ of the citizen. In the cloud-based approach, the MIS sends ($c_{CR}, \sigma_{MIS}$) to the CR. The CR verifies $\sigma_{MIS}$, decrypts $c_{CR}$, and searches its register for mandates using the plain $ssPIN_{CR}$. In our example, we assume that the mandate information $mand$ and the corresponding mandate ID $mandID$ has been found. The CR signs the mandate and the signature $\sigma_{CR}={\sf DSS.Sign}(sk_{CR}, mand\|mandID)$ is calculated.
	\item [8b.(new)] Since the citizen is known to the CR (the mandate contains further information of the citizen), it can encrypt the mandate for the citizen using $pk_{C}$ resulting in $c_{C}={\sf PK.Enc}(pk_{C}, mand\|mandID\|\sigma_{CR})$. The CR again signs the encryption result for ensuring integrity and authenticity calculating $\sigma_{CR}'={\sf DSS.Sign}(sk_{CR}, c_C)$.
	\item [8c.] In the current approach, the CR returns all registered mandate information for this citizen. In the cloud-based approach, the data ($c_C, \sigma_{CR}'$) are returned to the MIS, which verifies the signature\footnote{In our scenario, for simplicity the CR has been queried for mandate information only. However, the MIS actually queries all registers that have mandate information available.}.
	\item [9a.] The MIS presents the citizen a selection page of all available mandates for her. In the cloud-based example, $c_C$ is sent to the citizen.
	\item [9b.(new)] The citizen decrypts $c_C$ and verifies $\sigma_{CR}$.
	\item [9c.] The citizen selects the mandate she wants to use for authentication. In this scenario we assume that she wants to act on behalf of a company and thus selects $mandID$. In the cloud-based approach, the citizen signs $mandID$ resulting in $\sigma_{C}={\sf DSS.Sign}(sk_{C}, mandID)$. $(mandID, \sigma_{C})$ are returned to the MIS.
	\item [10.(n/a)] The MIS assembles all necessary mandate information and signs these data to generate an electronic mandate according to the specification defined by \cite{Rossler2006}. Amongst others, this electronic mandate contains information of the citizen, who represents the company, the company, and the type of empowerment the citizen is allowed to act on behalf. This step is not applicable in the cloud approach as the mandate information is a valuable asset which must not to be disclosed to the cloud provider.
	\item [11a.(new)] The MIS queries again the CR for retrieving all necessary information for the selected mandate by using $(mandID, \sigma_{C})$.
	\item [11b.(new)] The CR calculates $c_{MIS}={\sf RE.Enc}(MIS, mand\|mandID)$ and signs it resulting in $\sigma_{CR}''={\sf DSS.Sign}(sk_{CR}', c_{MIS})$.
	\item [11c.(new)] The CR transmits ($c_{MIS}, \sigma_{CR}''$) to the MIS. 
	\item [12.]  The MIS verifies $\sigma_{CR}''$ and re-encrypts $c_{MIS}$ for MOA-ID, i.e., computes $c_{\M}={\sf RE.ReEnc}(c_{MIS}, rk_{{MIS} \rightarrow \M})$. The MIS signs this re-encryption result $c_{\M}$ by calculating the signature $\sigma_{MIS}'={\sf DSS.Sign}(sk_{MIS}', c_{\M})$. The steps 11a.-12. are only required in the cloud-based approach.
	\item [13.] In the current approach, the MIS returns the electronic mandate to MOA-ID. In the cloud-based approach, the MIS returns ($c_{\M}, \sigma_{MIS}'$) to MOA-ID. MOA-ID verifies $\sigma_{MIS}'$.
	\item [14.(new)] MOA-ID re-encrypts the data $c_{\M}$, encrypted $ssPIN_{SP}$, and $c_{a_i}$ for $S_j$ using the key $rk_{\M \rightarrow S_j}$. The result $c_{S_j}$ is additionally signed using $sk_{\M}'$ resulting in $\sigma_{\M}'$. MOA-ID assembles ($c_{S_j}, \sigma_{\M}')$ in the SAML response.
	\item [15.] MOA-ID assembles an appropriate SAML assertion/response including the electronic mandate and transmits it to the online application.
	\item [16a.] The online application verifies the response. In the cloud-based approach, the online application verifies the signature $\sigma_{\M}'$.
	\item [16b.(new)] The online application decrypts the mandate and citizen information $c_{S_j}$ by using the key $sk_{S_j}$
	\item [17.] If verification is successful the online application grants access. The citizen is now able to do online procedures on behalf of the selected company.
\end{enumerate}

\subsubsection{Identification and Authentication of Foreign Citizens}
\label{ia_foreign_citizens}

In this section the identification and authentication of foreign citizen in the current approach and in the cloud approach are described in more detail.

\paragraph{Setup}

In the previous scenarios we assumed the SRA to be the trusted entity that issues appropriate key material to the involved entities in the Austrian eID system. In this scenario we have to deal with a cross-border scenario, hence we need a trusted entity being able to serve entities across borders. In the current STORK concept, the European Commission (EC) plays a central role managing trust across the involved STORK entities. Therefore, also for our scenario we assume the EC being the entity that issues secure key material to the individual STORK entities and thus we skip a detailed description on that. In this scenario, the EC generates $(pk_{PEPS}, sk_{PEPS})$ and issues $pk_{PEPS}$ to the PEPS only. It keeps secret $sk_{PEPS}$. Furthermore, the SRA issues $(pk_{MOA-ID}, sk_{MOA-ID})$, $(pk_{SPR-GW}, sk_{SPR-GW})$, $(pk_{SP}, sk_{SP})$, and $(pk_{SR}, sk_{SR})$. It keeps secret $sk_{MOA-ID}$ and $sk_{SPR-GW}$. The other keys are distributed to the respective entities. In addition, EC generates a re-encryption key $rk_{{PEPS} \rightarrow {MOA-ID}}$ and the SRA the re-encryption keys $rk_{{MOA-ID} \rightarrow {SPR-GW}}$, $rk_{{SPR-GW} \rightarrow {SR}}$, and $rk_{{MOA_ID} \rightarrow {SP}}$. For further explanations, we denote the identity data of the foreign citizen as ${fc}_{data}$.

We further assume that appropriate signing keys are available for the individual entities: $(pk_{\M}', sk_{\M}')$, $(pk_{PEPS}', sk_{PEPS}')$, $(pk_{F-IdP}', sk_{F-IdP}')$, $(pk_{SPR-GW}', sk_{SPR-GW}')$, and $(pk_{SR}', sk_{SR}')$.

\paragraph{Process Flow}

\begin{figure}[ht!]
	\centering
		\includegraphics[width=13cm]{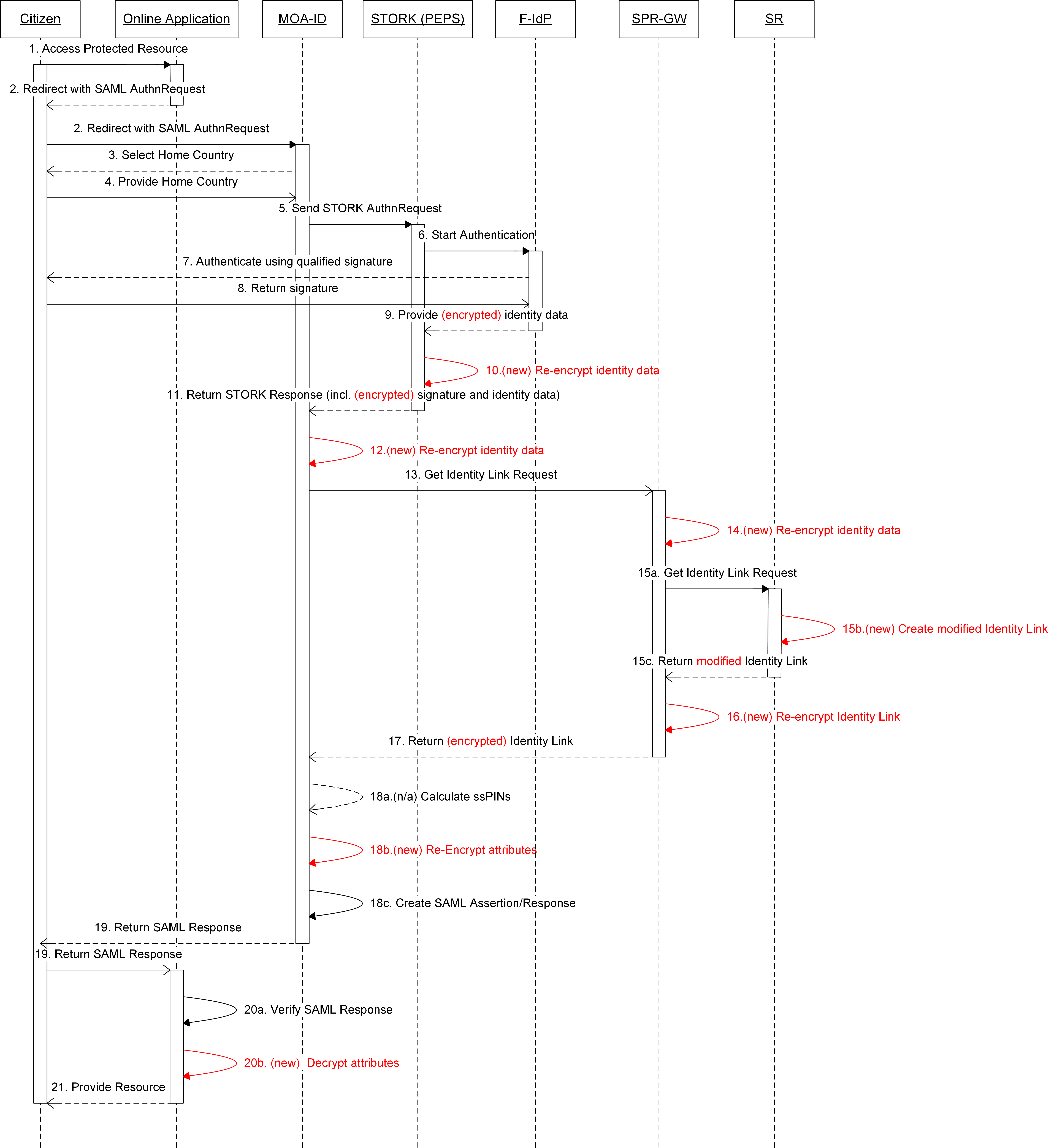}
	\caption{Process flow representing identifying and authenticating a foreign citizen in the cloud approach}
	\label{fig:seq_foreign_cloud}
\end{figure}

Figure \ref{fig:seq_foreign_cloud} illustrates the process flow identifying and authenticating a foreign citizen in the current and cloud-based approach. In the following, we describe the process flow in detail.

\begin{enumerate}
	\item [1.] A foreign EU citizen wants to access a service of an Austrian online application.
	\item [2.] The online application assembles an appropriate SAML authentication request and sends it to MOA-ID.
	\item [3.] MOA-ID presents the foreign citizen a page where the citizen can select her country of origin.
	\item [4.] The citizen provides her home country she originates from.
	\item [5.] According to the STORK idea, the foreign citizen will be authenticated in her home country. Therefore, the citizen is redirected to a single gateway (PEPS) in the foreign country, being part of the STORK infrastructure. For starting this authentication process, MOA-ID transmits a STORK authentication request to the foreign PEPS. The PEPS selects an appropriate foreign IdP (F-IdP), where the citizen actually authenticates.
	\item [6.] The PEPS forwards the authentication request to the F-IdP.
	\item [7.] The F-IdP requests the citizen to authenticate using a qualified signature.
	\item [8.] The qualified signature is returned to the F-IdP.	
	\item [9.] In the current approach, the F-IdP provides the qualified signature as well as other citizen identifying information (first name, last name, date of birth, identifier) to the PEPS. In the cloud-based approach, we assume the Foreign IdP to be a trusted entity and that it encrypts the foreign citizen's identification data for the PEPS using $pk_{PEPS}$ resulting in  $c_{PEPS}={\sf RE.Enc}(PEPS, {fc}_{data})$. Furthermore, $c_{PEPS}$ is signed using $sk'_{F-IdP}$ resulting in $\sigma_{F-IdP}$. Both results ($c_{PEPS}, \sigma_{F-IdP}$) are sent to the PEPS. The PEPS verifies $\sigma_{F-IdP}$.
	\item [10.(new)] The PEPS re-encrypts $c_{PEPS}$ for MOA-ID using $rk_{PEPS \rightarrow \M}$ into $c_{\M}$. In addition, $c_{\M}$ is signed using $sk'_{PEPS}$ resulting in $\sigma_{PEPS}$.
	\item [11.] In the current approach, the PEPS assembles the citizen data retrieved from the F-IdP and returns a STORK response to MOA-ID. In the cloud-based approach, the tuple ($c_{\M}, \sigma_{PEPS}$) is sent to MOA-ID. MOA-ID verifies $\sigma_{PEPS}$.
	\item [12.(new)] MOA-ID again re-encrypts the foreign citizen data for the SPR-GW to $c_{SPR-GW}={\sf RE.ReEnc}(rk_{\M \rightarrow SPR-GW}, c_{\M})$. $c_{SPR-GW}$ and the governmental sector $s$ of the SP is signed resulting in $\sigma_{MOA-ID}={\sf DSS.Sign}(sk_{\M}', c_{SPR-GW}\|s)$.
	\item [13.] MOA-ID extracts this information and sends it to the SPR-GW. The SPR-GW verifies the citizen's signature. In the cloud-based approach, the tuple ($c_{SPR-GW}, s, \sigma_{MOA-ID}$) is sent to the SPR-GW. The SPR-GW verifies $\sigma_{MOA-ID}$.
	\item [14.(new)] The SPR-GW does the re-encryption for the SR: $c_{SR}={\sf RE.ReEnc}(rk_{SPR_GW \rightarrow SR}, c_{SPR-GW})$. Again, the values $c_{SR}$ and $s$ are signed resulting in $\sigma_{SPR-GW}$.
	\item [15a.] The SPR-GW queries the SR to register the foreign citizen in the Supplementary Register for Natural Persons (SR) based on the information received. This registration into the SR is legally based on the Austrian e-Government act \cite{FederalChancellery2008} and the Austrian e-Government equivalence decree \cite{gleichwert_vo}. In the cloud-based approach, the tuple ($c_{SR}, s, \sigma_{SPR-GW}$) is sent to the SPR-GW. 	
	\item [15b.(new)] The SR verifies $\sigma_{SPR-GW}$, decrypts $c_{SR}$ using $sk_{SR}$, and registers the foreign citizen. During registration, a new modified Identity Link $\cal I'$ is created for the foreign citizen. Since the modified Identity Link is created on the fly, it just contains the encrypted $ssPIN$ for the sector $s$ and all other $ssPINs$ are redacted.
	\item [15c.] The SR calculates a sourcePIN for the citizen, creates and assembles an Identity Link, and returns the signed Identity Link to the SPR-GW. In the cloud-based approach, the new modified Identity Link $\cal I'$ is encrypted for the SPR-GW using $pk_{SPR-GW}$ resulting in $c'_{SPR-GW}$. The result $c'_{SPR-GW}$ is signed by applying $\sigma_{SR}={\sf DSS.Sign}(sk_{SR}', c'_{SPR-GW})$. Both results ($c'_{SPR-GW}, \sigma_{SR}$) are transferred to the SR. The SPR-GW again verifies $\sigma_{SR}$.
	\item [16.(new)] The SPR-GW again re-encrypts $c'_{SPR-GW}$ to $c'_{\M}={\sf RE.ReEnc}(rk_{SPR_{GW} \rightarrow \M}, c'_{SPR-GW})$, and signs the result $c'_{\M}$ applying $\sigma'_{SPR-GW}={\sf DSS.Sign}(sk_{SPR-GW}', c'_{\M})$.
	\item [17.] The SPR-GW returns the Identity Link to MOA-ID. In the cloud-based approach, the re-encrypted Identity Link $c'_{\M}$ and $\sigma'_{SPR-GW}$ are transmitted to \M. MOA-ID verifies $\sigma_{SPR-GW}$.
	\item [18a.(n/a)] MOA-ID derives the appropriate ssPIN out of the sourcePIN for the sector the online application belongs to. This step is not applicable in the cloud approach as the sourcePIN as well as the ssPIN are valuable assets which must not to be disclosed to the cloud provider.
	\item [18b.] MOA-ID re-encrypts to $c_{SP}={\sf RE.ReEnc}(rk_{\M \rightarrow SP}, c'_{\M})$, and signs $c_{SP}$ applying $\sigma'_{\M}={\sf DSS.Sign}(sk_{\M}', c_{SP})$. 
	\item [18c.] MOA-ID assembles a SAML assertion/response to be transferred to the SP. In the cloud-based approach the SAML response includes $c_{SP}$ and $\sigma'_{\M}$ .
	\item [19.-21.] These process steps are equal to normal and cloud-based Austrian citizen authentication (cf. Section \ref{ia_austrian_citizens}). In the cloud-based approach, the online application verifies $\sigma'_{\M}$ and decrypts $c_{SP}$ using $sk_{SP}$.	
\end{enumerate}

\section{Analysis and Discussion of the Proposed Model}
\label{sec:Discussion}

In this section we discuss the proposed migration of the Austrian eID system into the public cloud concerning security and privacy as wells practicability aspects.

\subsection{Security and Privacy Discussion}

Our work is based on the assumption that a cloud provider hosting or operating an entity is acting \emph{honest but curious} \cite{Chen2010a, Nunez2014}, i.e., the cloud provider operates and works correctly but is not trusted with respect to (data) privacy\footnote{A discussion of security and privacy issues when acting with a totally untrusted cloud provider is out of scope of this work and left for future work. However, under such an assumption data confidentiality and data integrity can still be ensured due to the use of encryption and signature technologies.}. In this section we investigate and discuss which personal and sensitive data are disclosed to an entity of the Austrian eID system operated in a public cloud. We thereby compare the information disclosed or seen, respectively, by an entity operated by a public cloud provider. Table \ref{tab:comparison} illustrates the comparison of the current Austrian eID system and the ported eID system to the cloud with respect to personal or sensitive data disclosed. Since encrypted data is seen as privacy-preserving data, any encrypted data disclosed at an entity in the cloud will not be mentioned.

\begin{table}[ht!]
  \centering
	\tiny
  \caption{Comparison of personal data disclosure between the current and the cloud-based approach}
		\begin{tabularx}{\textwidth}{|X|X|X|X|X|X|X|}
		\hline
		
    \multirow{2}[3]{2cm}{\textbf{Approach}} & \multirow{2}[3]{2cm}{\textbf{Use Case}} & \multicolumn{4}{>{\columncolor{grey}}c|}{\textbf{Component}} \\ \cline{3-6}
		
          &       & \multicolumn{1}{>{\columncolor{grey}}c|}{\emph{MOA-ID}} & \multicolumn{1}{>{\columncolor{grey}}c|}{\emph{MIS}} & \multicolumn{1}{>{\columncolor{grey}}c|}{\emph{SPR-GW}} & \multicolumn{1}{>{\columncolor{grey}}c|}{\emph{PEPS}} \\ \hline
		
    \multirow{3}[6]{2cm}{Current Approach} & Austrian Citizens & \multicolumn{1}{X|}{		
		\begin{itemize}[leftmargin=0.2mm,labelindent=0.2mm,labelsep=0.5mm, nolistsep]
  \item Identity Link (name, date of birth, sourcePIN)
  \item ssPIN
	\item Signing certificate
	\item Governmental sector
\end{itemize}
		} & -     & -     & - \\ \cline{2-6}
          & In Representation & \multicolumn{1}{X|}{
					\begin{itemize}[leftmargin=0.2mm,labelindent=0.2mm,labelsep=0.5mm, nolistsep]
  \item Identity Link (name, date of birth, sourcePIN)
  \item ssPIN
	\item Signing certificate
	\item All information of the mandate
	\item Selected mandate for application
	\item Governmental sector
\end{itemize}
} & \multicolumn{1}{X|}{
\begin{itemize}[leftmargin=0.2mm,labelindent=0.2mm,labelsep=0.5mm, nolistsep]
  \item Identity Link (name, date of birth, sourcePIN)
  \item ssPIN
	\item Signing certificate
	\item All registered mandate information of the citizen
	\item Selected mandate for application
	\item Governmental sector
\end{itemize}
} & -     & - \\ \cline{2-6}
          & Foreign Citizens & \multicolumn{1}{X|}{
					\begin{itemize}[leftmargin=0.2mm,labelindent=0.2mm,labelsep=0.5mm, nolistsep]
  \item Citizen's home country
  \item All requested citizen data (name, date of birth, identifier)
	\item Signing certificate
	\item Identity Link (sourcePIN, etc.)
	\item ssPIN
	\item Governmental sector
\end{itemize}
} & -     & \multicolumn{1}{X|}{
\begin{itemize}[leftmargin=0.2mm,labelindent=0.2mm,labelsep=0.5mm, nolistsep]
  \item Citizen's home country
  \item All requested citizen data (name, date of birth, identifier)
	\item Signing certificate
	\item Identity Link (sourcePIN, etc.)
\end{itemize}} & \multicolumn{1}{X|}{
\begin{itemize}[leftmargin=0.2mm,labelindent=0.2mm,labelsep=0.5mm, nolistsep]
  \item All requested citizen data (name, date of birth, identifier)
	\item Signing certificate
\end{itemize}} \\ \hline 
    \multirow{3}[6]{2cm}{Cloud-based Approach} & Austrian Citizens & \multicolumn{1}{X|}{
		\begin{itemize}[leftmargin=0.2mm,labelindent=0.2mm,labelsep=0.5mm, nolistsep]
  \item Governmental sector
\end{itemize}} & -     & -     & - \\ \cline{2-6}
          & In Representation & \multicolumn{1}{X|}{
					\begin{itemize}[leftmargin=0.2mm,labelindent=0.2mm,labelsep=0.5mm, nolistsep]
  \item Governmental sector
\end{itemize}} & 
\begin{itemize}[leftmargin=0.2mm,labelindent=0.2mm,labelsep=0.5mm, nolistsep]
  \item MandateID
\end{itemize} & -     & - \\ \cline{2-6}
          & Foreign Citizens & \multicolumn{1}{X|}{
					\begin{itemize}[leftmargin=0.2mm,labelindent=0.2mm,labelsep=0.5mm, nolistsep]
  \item Governmental sector
\end{itemize}} & -     & -     & - \\ \cline{2-6}
    \hline
    \end{tabularx}%
  \label{tab:comparison}%
\end{table}%

The comparison Table \ref{tab:comparison} follows the structure of the previous chapters and sections, where three different use cases for the Austrian eID system are distinguished (Identification and Authentication of citizens, in representation, of foreign citizens). In the following, privacy-sensitive data, which is revealed to the individual components, is discussed in detail. However, we only compare those components which are finally ported into the public cloud as they can be considered untrusted with respect to privacy. All other components are trusted and thus do not need a further analysis.

\textbf{Identification and Authentication of citizens:}

In this use case only MOA-ID is involved, hence only this component needs to be investigated with respect to privacy. In the current scenario the citizen's identity link (including her name, date of birth and sourcePIN) is exposed to MOA-ID. Additionally, MOA-ID knows the governmental sector of the application the user wants to authenticate and thus also the ssPIN, which is derived out of the citizen's sourcePIN by MOA-ID. Finally, also the citizen's signing certificate is disclosed to MOA-ID.

In contrast to this data set, in the cloud-based approach only the governmental sector of the application the citizen wants to log in remains visible to MOA-ID and the cloud provider respectively. All other data is transferred in encrypted form to MOA-ID only.

\textbf{Identification and Authentication in representation:}

In this scenario the components MOA-ID and MIS are involved. Equal of the previous use case, in the current approach MOA-ID gets to know the citizen's identity link (including her name, date of birth and sourcePIN), the citizen's signing certificate, and subsequently the governmental sector of the application and the citizen's ssPIN. The identity link data is also disclosed to the MIS. Since the MIS handles all relevant functionality with respect to authentication on behalf, the MIS also sees all all registered mandate information of the citizen. The reason is that the MIS queries all available registers to find existing mandate information for the authenticating citizen. This mandate information is bundled at the MIS for displaying it to the user. After selecting a mandate by the user, the MIS also knows which mandate has been selected. Due to that, the MIS also gets the information of the empowering mandator. Since MOA-ID and the MIS are interconnected, for fulfilling a successful authentication process in representation all selected mandate information (mandate type, mandator, empowerment, etc.) is also disclosed to MOA-ID. Hence, a lot personal information is disclosed to both components.

Having a look at the cloud approach again, only a minimum of those data are disclosed to MOA-ID and the MIS in this setup. During an authentication process in representation, the MIS only learns the ID (mandID) of the selected mandate by the citizen. In addition, MOA-ID only gets to know the governmental sector of the application. No further privacy-sensitive data is disclosed to either MOA-ID or the MIS because it is processed in encrypted form only.

\textbf{Identification and Authentication of foreign citizens:}

In this use case the components MOA-ID, SPR-GW, and PEPS are involved. Already at the start of an authentication request, MOA-ID gets to know the citizen's home country because MOA-ID needs to forward the citizen to the respective PEPS for authentication. After successful authentication of the foreign citizen at the F-IdP, all requested citizen data (name, date of birth, unique identifier, signing certificate)\footnote{The STORK framework and protocol also supports the transfer of many more attributes. However, for the authentication at an Austrian service provider the name, date of birth, unique identifier, and signing certificate are sufficient, hence we skip a detailed discussion of additional attributes.} is disclosed to the PEPS for further processing. All these data is returned from the PEPS to MOA-ID, thus these data is also disclosed MOA-ID. For registering the foreign citizen in the Austrian Supplementary Register of Residents, these data are forwarded by MOA-ID to the SPR-GW. Out of the signing certificate, the SPR-GW also gets to know the citizen's home country. The SPR-GW registers the citizen and as return data it receives the Austrian identity link of the foreign citizen. This identity link also includes the Austrian unique identifier (sourcePIN), which constitutes sensitive information. The identity link is further exposed to MOA-ID, which uses the governmental sector and the sourcePIN for calculation of the ssPIN.

Again, compared to all exposed information described above in the current setting, only the governmental sector of the application is disclosed to MOA-ID in the cloud approach. No further sensitive information is disclosed to any of the components MOA-ID, SPR-GW, or PEPS. All data is only available in encrypted form at these components.

\subsection{Practicability Discussion}

In this section the proposed cloud approach based on selected criteria with respect to to practicability is discussed. The following criteria were selected:

\begin{description}
	\item[Re-Use of Existing Infrastructure:] 
	When designing the proposed solution, one criterion was that the existing infrastructure should not severely altered or changed. This particularly includes the architecture and the functionality of the individual components. When comparing the Figures \ref{fig:eID-Architecture} and \ref{fig:eID-Architecture-Cloud} it can be seen that the overall architecture remains the same instead of the migration of MOA-ID, MIS, SPR-GW, and PEPS into a public cloud. Even the message and transport protocols used to exchange messages between the individual components do not require heavy changes. The exchange messages and protocols need to support the transfer of encrypted data in the cloud-based approach. In addition, also trusted components need to be made capable of encryption functionality. However, this keeps the effort to a minimum, hence no severe changes to the Austrian eID infrastructure need are made.
	\item[Conformance to Current Process Flow:]
	Another design criterion for our approach was staying conform to the current process flow. While in general conformity is mostly given,  changes in the process flow are required to keep the high level of security and privacy-preservation with respect to the cloud providers. However, those process flow changes can be seen as minimal as they are only related to encryption, re-encryption, or additional signature steps.
	\item[Scalability:]
	The main aim of this work was to guarantee high scalability for central services even at a huge number of users. This aim is mainly realized by migrating important components such as MOA-ID, SPR-GW, or the MIS, where high load can be expected, into public clouds. Of course, load bottlenecks might occur at trusted message endpoints such as individual registers or the F-IdP for the foreign citizen use case. However, those situations are probably very unlikely because not for all citizens the same registers need to be queried at the same time or foreign citizens will not use the same F-IdP at the same time.
	\item[Governance Structure:]
	For the current situation there is a proper governance structure in place. Meaning, proper trust relationships are (e.g. based on digital certificates) between the individual components. However, the use of re-encryption functionality adds an additional layer of governance requirements. Encryption and decryption key pairs as well as re-encryption keys need to be properly distributed amongst the involved components. This puts some additional complexity to the SRA for use cases involving Austrian citizens only. In addition, also the European Commission and foreign countries are affected as re-encryption functionality is also required across borders. In particular, key management and distribution based on a public key infrastructure (PKI) to the individual components needs to be carried out properly. However, the effort for these tasks can be considered reasonable as the number of involved components is limited (In Austria besides MOA-ID, SPR-GW, and the MIS a few registers and several service providers, in foreign countries besides the PEPS especially different F-IdPs).
\end{description}

\subsection{Related Work Discussion}

\begin{description}
	\item[Related Work on using Proxy Re-Encryption:] One of the first approaches using proxy re-encryption for identity management in the cloud appeared in \cite{Nunez2012}, which integrated proxy re-encryption into the OpenID protocol. They continued their work by creating a generalized model called \emph{BlindIdM} (A Privacy-Preserving Approach for Identity Management as a Service) \cite{Nunez2014}. The applicability of \emph{BlindIdM} was further demonstrated by integrating proxy re-encryption functionality into SAML. However, \cite{Nunez2012,Nunez2014} rely on proxy re-encryption for privacy preservation only, whereas our approach additionally uses on redactable signatures, as an additional privacy-preserving mechanism ensuring integrity and authenticity at the same time.
	\item[Related Work on using Anonymous Credentials:] Anonymous Credential systems (aka Privacy ABCs) are a valuable mechanism for ensuring privacy in identity management and have been discussed in context of eID systems, e.g, their integration in the German eID architecture \cite{Bjones2014}. While early implementations of anonymous credential systems, e.g., idemix \cite{Zurich2010} on a Java Card \cite{Bichsel2009}, however, were too expensive from the user's (client's) perspective, state-of-the-art implementations~\cite{DBLP:conf/cans/PiedraHV14,Vullers2014,DBLP:conf/rfidsec/HinterwalderRP15} already achieve reasonable efficiency. While anonymous credentials are a valuable means for ensuring privacy in identity management, performance is typically still much slower than when using proxy re-encryption.
\end{description}

\section{Conclusions}
\label{sec:Conclusions}

The Austrian eID system plays a major role in the Austrian e-Government strategy. Its main functions are unique identification and secure authentication of Austrian citizens, in representation on behalf of a natural or legal person, or foreign citizen authentication. The current Austrian eID system is based on several components, which are interconnected amongst others. Some of the components are deployed locally (MOA-ID), others centrally (MIS, SPR-GW, PEPS). In general, a central deployment of each individual component is preferable. For instance, a central deployment saves service providers a lot of operational and maintenance costs. However, a central deployment can easily lead to load bottlenecks and scalability issues when the frequency of identification and authentication processes increases. Theoretically, the entire population of Austria and -- going beyond boarders -- of whole Europe will be able to use and run authentications through the Austrian eID system.

To overcome such scalability bottlenecks in the future, in this paper we presented a solution by moving important centralized services of the Austrian eID system (MOA-ID, MIS, SPR-GW, and PEPS) into a public cloud which considerably improves scalability. By applying appropriate cryptographic technologies we are able to improve scalability by preserving citizen's privacy with respect to the public cloud providers at the same time. We therefore can conclude that for all identification and authentication use cases no sensitive personal information will be disclosed to a public cloud provider in the cloud-based approach since all data processed in the cloud is encrypted. This strongly preserves citizen's privacy even if public cloud providers assuming to be acting honest but curious are involved in the proposed architecture. In addition, no major changes to existing infrastructure or the current process flows are required. Only decryption/encryption/re-encryption functionality need to be additionally supported by the individual components and the data transfer protocols must be capable of encrypted data. However, efforts for implementing encryption functionality might be low and encrypted data can also easily transmitted by standard data exchange protocols such as SAML \cite{Nunez2014} or OpenID \cite{Nunez2012}. Finally, a proper governance structure for additional management of encryption/decryption/re-encryption keys needs to be setup. Nevertheless, this can be easily integrated into existing organizational procedures.

\section*{References}

\bibliography{JISA2015}

\end{document}